# Emergence of Nontrivial Topological Magnon States in Skyrmionium Lattices with Zero Topological Charge

Xingen Zheng[1,2], Ping Tang[3], Xuejuan Liu[4], Zhixiong Li[5], Peng Yan[6*], and Hao Wu[1,2*]

[1]Dongguan Institute of Materials Science and Technology, Chinese Academy of Sciences, Guangdong 523808, China

[2]Songshan Lake Materials Laboratory, 523808, Dongguan, Guangdong, China

[3] Institute for Materials Research, Tohoku University, 2-1-1 Katahira, Sendai 980-8577, Japan

[4]Shenzhen Key Laboratory of Ultraintense Laser and Advanced Material Technology, Center for Intense Laser Application Technology, and College of Engineering Physics, Shenzhen Technology University, Shenzhen 518118, China

[5]School of Physics, Central South University, Changsha 410083, China

[6]State Key Laboratory of Electronic Thin Film and Integrated Devices, School of Physics, University of Electronic Science and Technology of China, Chengdu 611731, China

[*]Author to whom any correspondence should be addressed. E-mail: wuhao1@sslab.org.cn, yan@uestc.edu.cn

## Abstract

**We predict the emergence of nontrivial topological magnon states in the skyrmionium lattice with zero topological charge. We propose the concept of weighted magnetic flux, which provides a clear physical picture for this anomalous phenomenon. We also map the skyrmionium lattice onto the Haldane model, offering an alternative framework for interpreting this. Our findings challenge the conventional wisdom that such states are linked to nonzero topological charge in skyrmion lattices, offering a new perspective in topological magnonics. To facilitate experimental validation, we propose two methods for preparing the skyrmionium lattice and calculate the induced magnon thermal Hall conductivity—a key indicator in transport measurements.**

Topological insulators are quantum materials with conducting edge/surface states but an insulating bulk, representing a frontier in condensed matter physics. Research on topological phenomena has expanded beyond electrons to photons, phonons and magnons [1-5]. Among these, topological magnons have attracted significant attention due to their absence of Joule heating and low dissipation in long-distance transport, leading to discoveries of topological magnon insulators [6-13] and magnon thermal Hall effect [14-20]. Skyrmions are topologically protected magnetic textures with nonzero topological charge $Q = \frac{1}{4\pi}\int \rho_{top} d^2r$, with topological charge density $\rho_{top} = \boldsymbol{m} \cdot (\partial_x \boldsymbol{m} \times \partial_y \boldsymbol{m})$. They tend to form periodic patterns known as **the skyrmion lattice (SkL)** [21]. Nontrivial topological magnon states can arise spontaneously in SkL from the induced **emergent magnetic field (EMF)** $\boldsymbol{B}_{em} = B_{em}\hat{z} = \frac{4\pi\hbar}{e}\rho_{top}\hat{z}$, where $\hat{z}$ is the normal of the SkL plane [22], resulting in topological magnon states and thermal Hall effect [23].

Then, a fundamental question arises: Is $Q \neq 0$ necessary for topological magnon states in SkLs? One might speculate that antiferromagnetic SkL may possess zero topological charge, leading to the absence of topological magnon states. While at least in the antiferromagnetic SkLs, the very existence of such topological states has been claimed [24]. However, this three-sublattice antiferromagnetic SkL remains to be proven $Q \neq 0$. To date, in nearly all previous studies of topological magnon states in skyrmion-based systems, both theoretically and experimentally, it has been implicitly or explicitly assumed $Q \neq 0$ is necessary for topological magnon bands [20,22-24]. This assumption stems from the idea that the quantized net emergent magnetic field is proportional to $Q$, leaving the aforementioned question unsolved.

In this Letter, we study **the skyrmionium lattice (SkML)**, a periodic arrangement of $Q$=0 skyrmioniums, which can be viewed as a composite structure formed by nesting two concentric skyrmions with opposite skyrmion charges [21,25]. By calculating the Chern number of the magnon band, we demonstrate, for the first time, the existence of topological magnon edge states in SkML with zero topological charge. To explain this, we propose a physical picture and map the SkML to the Haldane model. We also derive the magnon thermal Hall conductivity as a measurable signature for experiments.

*Model and magnon band structure.* We consider a triangular spin-lattice Hamiltonian

$$H = -\sum_{<i,j>} J_{ij}\vec{S}_i \cdot \vec{S}_j + \sum_{<i,j>} \vec{D}_{ij} \cdot (\vec{S}_i \times \vec{S}_j) - h_z \sum_i S_i^z - K \sum_i (S_i^z)^2, \quad (1)$$

with the spin-$S$ operator $\vec{S}_r$ at site $r$. Here $J_{ij}$, $D_{ij}$, $K$, and $h_z$ represent the nearest-neighbor exchange, Dzyaloshinskii-Moriya interaction (DMI), single-ion uniaxial anisotropy, and applied magnetic field along *z*-direction, respectively. We consider Bloch-type DMI with $\boldsymbol{D} = D\boldsymbol{n}_{ij}$, where $\boldsymbol{n}_{ij}$ is the unit vector joining sites *i* and *j*. With the above model, parameters are taken in units of *J* as $D=\tan(\pi/8)J$, $K=0.15J$, $h_z=0.04J$ and the spin $S=1$ in the following calculations if not stated otherwise.

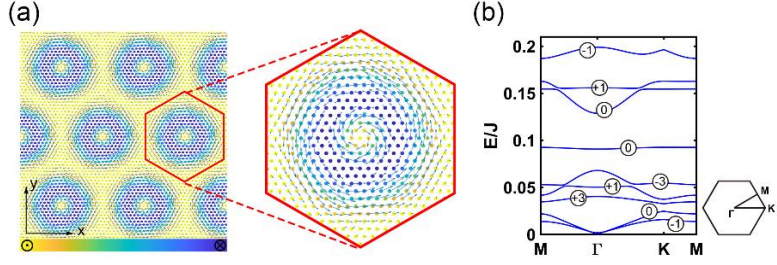

FIG. 1. (a) Left panel: Diagram of SkML. Right panel: Enlarged view of a magnetic unit cell. (b) Magnon band structure and corresponding Chern numbers for SkML. The momenta are selected along the high-symmetry points of the Brillouin zone (bottom-right corner). The Chern number for each band is indicated by an encircled number.

The phase diagram of this Hamiltonian typically includes three states: helical, **ferromagnetic (FM)**, and SkL states, depending on the temperature and magnetic field [26]. Although experimental methods for preparing SkML are currently lacking and only individual skyrmionium creation has been proposed [25,27-29], in theoretical calculations, a perfect SkML can be realized by specifying spin configurations and using the Monte Carlo simulation to relax the system, as shown in Fig. 1(a). The yellow-blue colormap represents the *z*-component of spin, and arrows show the in-plane components. The magnetic cell (22×22 spins) is outlined in red, with a zoom-in view on the right panel. To guide future experimental verification, we provide two possible methods for experimentally preparing SkML in **Supplemental Material S1** [30]. We also show in **Supplemental Material S2** how *Q*=0 is obtained for the SkML, confirming its topologically trivial nature.

With the Hamiltonian (1) of SkML, the magnon band structure and spatial distribution of the magnon wavefunction under low-energy excitation can be derived via the Holstein-Primakoff (HP) boson theory [5,23,52,53] and para-unitary transformation [54-56]. This procedure accesses the magnon bands $E(\boldsymbol{k})$ and eigenstates $\psi(\boldsymbol{k})$. Details of this procedure are provided in **Supplemental Material S3**. We display the lowest ten bands of the magnon band structure in Fig. 1(b), as we focus on low-energy excitations, with momenta selected along high-symmetry points. Since skyrmioniums demonstrate Newtonian dynamics and possess mass [57], two translational modes with linear $k$-dependence appear in the vicinity of the Γ-point.

**The Chern number and the magnon edge states.** To confirm the existence of topologically nontrivial magnon bands in SkML, we calculate the Chern number for each band using

$$C_j = \frac{(-1)^{\sigma_j}}{2\pi} \int_{BZ} \Omega_j(\boldsymbol{k}) d^2\boldsymbol{k}, \tag{2}$$

where $j$ labels the $j$-th magnon band; $\sigma_j = 0(1)$ for positive (negative) energy band; $\Omega_j(\boldsymbol{k})$ is the Berry curvature, given by

$$\Omega_j(k_x, k_y) = \lim_{dk_x, dk_y \to 0} \frac{-i \times \log(U_1 U_2 U_3 U_4)}{dk_x dk_y}, \tag{3}$$

where

$$U_1 = \langle \psi_j(k_x, k_y) | \sigma_3 | \psi_j(k_x + dk_x, k_y) \rangle,$$
$$U_2 = \langle \psi_j(k_x + dk_x, k_y) | \sigma_3 | \psi_j(k_x + dk_x, k_y + dk_y) \rangle,$$
$$U_3 = \langle \psi_j(k_x + dk_x, k_y + dk_y) | \sigma_3 | \psi_j(k_x, k_y + dk_y) \rangle,$$
$$U_4 = \langle \psi_j(k_x, k_y + dk_y) | \sigma_3 | \psi_j(k_x, k_y) \rangle. \tag{4}$$

Here, $dk_x(dk_y)$ is the infinitesimal of the $k_x(k_y)$, $\sigma_3$ is the direct product of the third Pauli matrix and a $N$-dimensional unit matrix. The Chern number for each band is indicated as an encircled number in Fig. 1(b), revealing that some bands exhibit a nonzero Chern number despite the skyrmionium having zero topological charge.

According to the bulk-edge correspondence [23,58], if the sum of Chern numbers for the lowest $n$ bands $C_n^{sum}$ is nonzero, $C_n^{sum}$ edge states appear between the $n$-th and $(n+1)$-th bands. To further demonstrate the edge states, we constructed a one-dimensional strip model by opening the boundary in the $y$-direction while maintaining periodicity in the $x$-direction, as the cylinder shown in the inset of Fig. 2(a). The magnon dispersion as a function of momentum $k_x$ for this strip is plotted in Fig. 2(a), with 22×161 spins, including 7 complete skyrmionium cells in the $y$ direction. In Fig. 2(a), each band in Fig. 1(b) splits into 7 sub-bands, and in-gap edge states appear above the bands with a non-zero $C_n^{sum}$. Here we select points "a" and "b" ("c" and "d") on the in-gap edge states, corresponding to $C_3^{sum} = 1$ ($C_2^{sum} = -1$), as marked by red and green dots in Fig. 2(a). The spatial distributions of magnon density at these points are shown in Fig. 2(b), where the distributions cluster at the bottom/top edges. Here for the given magnon state $|\psi\rangle$, the magnon density at site $i$ is calculated by $\rho_i \equiv \langle\psi|a_i^\dagger a_i|\psi\rangle$ [23,30]. Due to the opposite sign of $C_2^{sum} = -1$ and $C_8^{sum} = 1$, edge states with same group velocity $\partial E/\partial k_x$, such as dots "a(b)" and "c(d)", cluster at different edges.

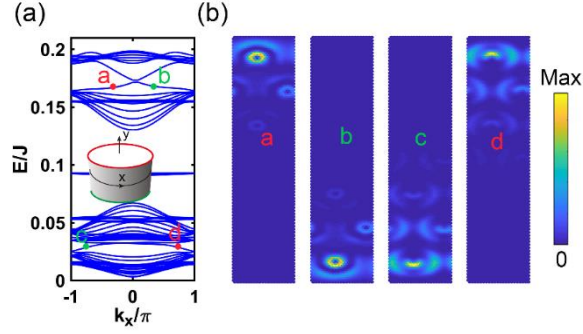

FIG. 2. Magnon edge states of SkML. (a) Magnon band structure for the one-dimensional strip of 7-skyrmioniums wide. (b) Magnon spatial distributions for four specific edge states corresponding to the red and green dots "a, b, c, d" in the band gap in (a).

*Physical picture for the nontrivial topological states.* This anomalous topological magnon state in SkML originates from the momentum space topology, i.e., the skyrmionium cell periodicity, which provides the magnon with such a possibility: When magnon at certain energy level propagates within the periodic lattice, its motion is mainly confined to specific region, such as the inner (outer) skyrmion within the skyrmionium, so that magnon mainly experience a EMF of the inner (outer) skyrmion, leading to its deflection.

To illustrate this, we show the magnon spatial distributions of the lowest eight energy levels within a unit cell under periodic boundary conditions in real space [see Figs. 3(a)-3(h)], corresponding to the lowest eight bands [Fig. 1(b)]. The dashed circles separate the inner and outer skyrmions within one skyrmionium. Qualitatively, for instance, magnon at the 1st/3rd energy level primarily localizes in the inner/outer skyrmion, indicating stronger influence from the EMF of the inner/outer skyrmion, corresponding to Chern number *C*<0/*C*>0. To quantitatively measure the effect of EMF for a magnon state, we consider both the magnon density distribution [Figs. 3(a)-3(h)] and strength of EMF distribution [Fig. 3(i)] in one cell.

We define the *weighted magnetic flux*

$$f = A_{uc} \int_{uc} \rho_{mag} B_{em} d^2 r, \tag{5}$$

where $A_{uc}$ is the area of the skyrmionium unit cell. The product of the magnon density $\rho_{mag}$ and the EMF $B_{em} = \frac{4\pi\hbar}{e}\rho_{top}$ (for convenience, $e = \hbar = 1$) is integrated over the unit cell. The physical meaning of $f$ is to capture the magnetic flux weighted by the magnon density for a certain magnon state. If $f>0$ ($f<0$), the EMF of the inner (outer) skyrmion contributes more for that magnon mode. For most cases, if $C \neq 0$, $C$ has an opposite sign with $f$. This relationship holds for all lowest eight energy levels as shown in Figs. 3(a)-3(h), where $C$ and $f$ are labeled for each diagram.

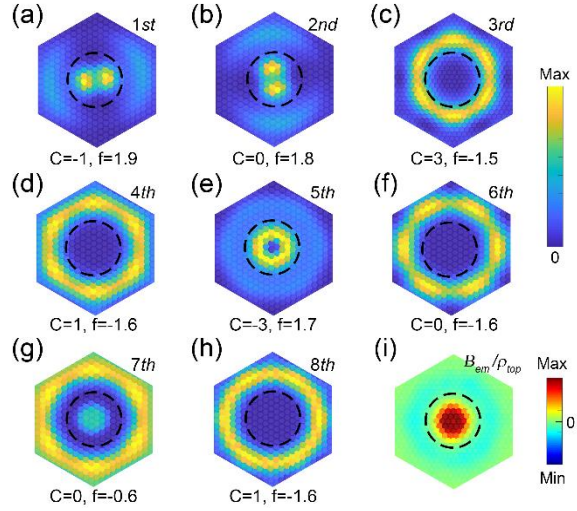

FIG. 3. (a)-(h) Magnon spatial distribution of the 1st-8th energy levels in one skyrmionium cell. The dashed circle separates the inner and outer skyrmions within the skyrmionium. The Chern number C and the weighted magnetic flux $f$ are labeled for each diagram. (i) The EMF distribution in one skyrmionium cell.

This approach provides an intuitive picture of how a $Q=0$ SkML can exhibit nontrivial topological states. Although it can determine the sign of the nonzero Chern number by assessing which EMF contributes more for the magnon and works for most bands, it is not universally applicable for all bands. This limitation arises because the EMF alone doesn't fully govern magnon dynamics—on-site potentials and intersite couplings also play a crucial role. If certain sites possess large on-site potentials and weak intersite couplings, the EMF's influence on these sites is suppressed. For instance, in the 6th band, the magnon is trapped by the strong on-site potentials and weak intersite couplings, rendering the band completely flat with $C=0$.

***Connection between skyrmionium lattice and Haldane model.*** We note that the magnon topology in SkML closely resembles that of the Haldane model [59], since both systems exhibit vanishing global magnetic flux, but with local flux split into two opposing parts.

As shown in Fig. 4(a), each unit cell in SkML, outlined by a red hexagon, is divided into six regions (black triangles), each shared by three neighboring cells. By treating all spins within each black triangle into one, the original unit cell simplifies into a hexagonal cell comprising six sites (A-F), as illustrated in Fig. 4(b). Nearest-neighbor, next-nearest-neighbor, and third-nearest-neighbor couplings between the six sites (A-F), represented by black, blue, and red lines, respectively. Here we denote hopping from site $\alpha$ to $\beta$ by $t_{\alpha\beta}$, with $t_{\alpha\beta}=t_{\beta\alpha}^{*}$.

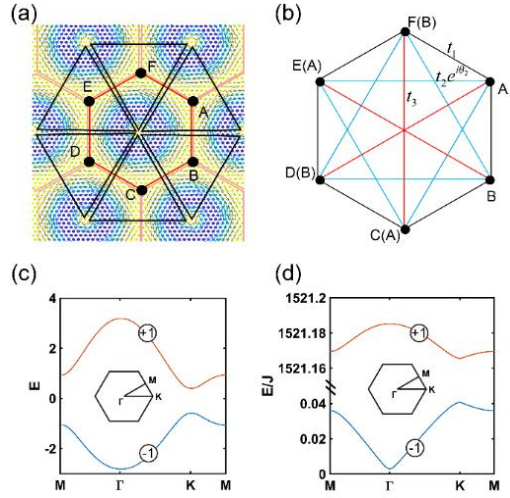

FIG. 4. (a) The skyrmionium cell (red hexagon) is approximated as a hexagonal cell with six sites (A-F) by grouping sites within black triangles. (b) The hexagonal lattice simplifies to the Haldane model when neglecting the weakest third-nearest-neighbor coupling $t_3$. (c) Band structure of the Haldane model for $t_1=1$, $t_2=0.2$, and $\theta_2=-0.4\pi$. Chern numbers are labeled in circles. (d) Two bands from SkML when summing the lowest two bands and the remaining bands, respectively. The corresponding summed Chern numbers are labeled in circles.

The hexagonal cell inherits the $C_6$ symmetry from SkML, so: (1) The six sites (A-F) share identical onsite potentials, giving an overall energy shift that can be subtracted. (2) Nearest-neighbor couplings $t_{AB}=t_{BC}=t_{CD}=t_{DE}=t_{EF}=t_{FA}$, represented as $t_1 e^{i\theta_1}$. Given that the zero topological charge as well as the associated zero total emergent magnetic flux needs $\theta_1 = 0$, the nearest-neighbor coupling is purely real $t_1$. (3) Next-nearest-neighbor couplings $t_{AC}=t_{CE}=t_{EA}=t_{BD}=t_{DF}=t_{FB}$, denoted as $t_2 e^{i\theta_2}$. $t_2$ is weaker than $t_1$ due to the longer distance. $\theta_2$ dictated by the EMF distribution, remains indeterminate. (4) Third-nearest-neighbor couplings $t_{AD}=t_{BE}=t_{CF}=t_{DA}=t_{EB}=t_{FC}$. Since $t_{\alpha\beta}=t_{\alpha\beta}^*$, the third-nearest-neighbor is real $t_3$, which is the weakest coupling with the longest distance. When neglecting the weakest $t_3$, this hexagonal lattice recovers the Haldane model. The six sites (*A-F*) are reduced to two effective sites (*A, B*), as labeled in Fig. 4(b).

We also find a connection in band structure between SkML and the Haldane model. For Haldane model, when $t_1=1$, $t_2=0.2$, $\theta_2=-0.4\pi$, the band structure is solved as shown in Fig. 4(c). Calculation details are provided in **Supplemental Material S4.** When $-\pi<\theta_2<0$, the lower (upper) band carries a Chern number $C=-1(1)$. The nonzero Chern numbers associated with the bulk bands indicate that topological edge states would necessarily emerge if a boundary were introduced. To unify the two models from the band structure, a reasonable picture is that when SkML is mapped into the Haldane model, the *N* bands of SkML merge into two bands of the Haldane model. Therefore, for SkML, we sum the lowest two bands and the remaining bands, separately. Then two bands, as illustrated in Fig. 4(d), are obtained, with the summed Chern numbers denoted on each band. Obviously, this band structure in Fig. 4(d) closely resembles that of the Haldane model in both shape and Chern numbers. From this view, the band structure and the nontrivial topological states in the Haldane model are inherited from SkML, which confirms our result that nontrivial topological states exist in SkML despite $Q=0$.

***The phase diagram and the thermal Hall conductivity.*** To investigate the thermal stability of SkML, we present a phase diagram [Fig. 5(a)] under varying temperatures and magnetic fields. Unlike traditional phase diagrams using random or FM states as the initial state followed by Monte Carlo annealing [26], we directly set SkML as the initial state [marked by a red star in Fig. 5(a)] and change the temperature and magnetic field to determine its final state. This phase diagram confirms that SkML is stable for lower magnetic fields at low temperatures.

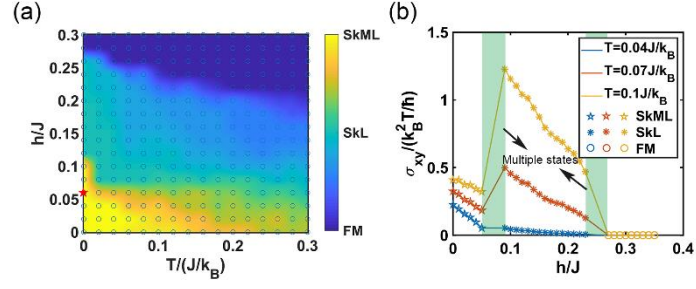

FIG. 5. (a) Phase diagram for SkML. The red star represents the initial state. (b) Thermal Hall conductivity for SkML, SkL, and FM as a function of temperature and magnetic field.

Furthermore, considering one practical way to verify the topological magnon state is by measuring the thermal Hall effect [20], we calculate the thermal Hall conductivity of SkML. We first obtain the magnetic structure at its minimum energy by relaxing the specific magnetic state at zero temperature, based on which the thermal Hall conductivity is evaluated by

$$\sigma_{xy} = -\frac{k_B^2 T}{\hbar} \int_{Bz} \frac{d^2 \mathbf{k}}{(2\pi)^2} \sum_{j=1}^{N} W\left[n_B(\varepsilon_j(\mathbf{k}))\right] \Omega_j(\mathbf{k}), \qquad (6)$$

where $\Omega_j(\mathbf{k})$ is the Berry curvature of the $j$th band, $W[x] = \int_0^x dt [\{(1+t)/t\}]^2$, and $n_B(\varepsilon) = 1/\{exp(\varepsilon/k_B T) - 1\}$ is the Bose distribution function [14,20]. The integration is performed over the first Brillouin zone, and the summation covers all $N$ bands. Since $\sigma_{xy}$ is determined by non-integer Berry curvature integration weighted by the Bose distribution, it could vary smoothly with field. We calculate $\sigma_{xy}$ for a series of magnetic structures by systematically varying the magnetic field at selected temperatures. Here we only select the three temperatures ($k_B T$=0.04, 0.07, 0.1$J$) whose phase boundary remains close in field and allows for a clean comparative analysis between the SkML and the SkL. Given the system near the phase transition is actually in multiple states and couldn't be well determined, $\sigma_{xy}$ near that region is omitted. The calculated field dependence of $\sigma_{xy}$ is shown in Fig. 5(b), revealing three key features: (1) In both SkML and SkL, $\sigma_{xy}$ decreases with increasing magnetic field. (2) The overall $\sigma_{xy}$ in SkML is lower than that in SkL. (3) In SkL, $\sigma_{xy}$ decreases sharply with decreasing temperature, while in SkML, the decrease is smoother. These features can be traced from the band structures. We give the analysis in **Supplemental Material S5**.

***Discussion and conclusion.*** In summary, we theoretically predicted the existence of nontrivial topological magnon bands and edge states in SkML, despite its zero topological charge. To explain this phenomenon, we propose a physical picture and build a connection between SkML and the Haldane model. For experimental verification, we calculated the thermal Hall conductivity as a key indicator of topological magnon edge states. While our study focuses on the triangular lattice and Bloch-type DMI, the findings are generalizable to square lattices and Néel-type DMI. Our work demonstrates the realization of nontrivial topological magnon bands and transport in SkML, paving the way for potential applications in magnonic logic and computing.

We thank Zhejunyu Jin for discussion. This work was supported by the National Key R&D Program of China (Grants No. 2022YFA1402801 and No. 2022YFA1402802), the National Natural Science Foundation of China (NSFC, Grant Nos. 52271239, 52311530674, 12434003, 12374103, and 12074057), the Guangdong Basic and Applied Basic Research Foundation (Grant Nos. 2022B1515120058, 2023A1515110880 and 2024A1515110196), Guangdong Provincial Quantum Science Strategic Initiative (GDZX2302003 and GDZX2301002), JSPS KAKENHI Grant for Early-Career Scientists (No. 23K13050).

# Supplemental Material for

# "Emergence of Nontrivial Topological Magnon States in Skyrmionium Lattices with Zero Topological Charge"

Xingen Zheng[1,2], Ping Tang[3], Xuejuan Liu[4], Zhixiong Li[5], Peng Yan[6*], and Hao Wu[1,2*]

[1]Dongguan Institute of Materials Science and Technology, Chinese Academy of Sciences, Guangdong 523808, China

[2]Songshan Lake Materials Laboratory, 523808, Dongguan, Guangdong, China

[3]Institute for Materials Research, Tohoku University, 2-1-1 Katahira, Sendai 980-8577, Japan

[4]Shenzhen Key Laboratory of Ultraintense Laser and Advanced Material Technology, Center for Intense Laser Application Technology, and College of Engineering Physics, Shenzhen Technology University, Shenzhen 518118, China

[5]School of Physics, Central South University, Changsha 410083, China

[6]State Key Laboratory of Electronic Thin Film and Integrated Devices, School of Physics, University of Electronic Science and Technology of China, Chengdu 611731, China

*Author to whom any correspondence should be addressed. E-mail: wuhao1@sslab.org.cn, yan@uestc.edu.cn

**S1 Two methods for preparing the skyrmionium lattice in experiment**

In this supplementary section, we provide two possible methods for the experimental preparation of the skyrmionium lattice (SkML). In both our methods, we create an environment conducive to SkML formation based on skyrmion lattice (SkL), then induce the SkL to transform into a SkML. And we employ the Monte Carlo simulation to model spin dynamics.

It is important to note that the mainstream techniques for creating an individual skyrmionium structure rely on starting from a pre-existing skyrmion. These methods generally fall into two categories: (1) nucleating a second, oppositely polarized core at the center of a skyrmion, thereby converting it into a skyrmionium [1-3], or (2) adding an extra 360° domain wall around the perimeter of a skyrmion to form a skyrmionium [4-5].

To date, no one has experimentally produced a SkML. This is because standard field-cooling or annealing procedures naturally drive the system into the SkL, helical, or ferromagnetic phases. In other words, the SkML corresponds to a metastable state that is essentially invisible in conventional equilibrium phase diagrams and can only be realized experimentally via externally assisted induction. However, based on these two mainstream approaches for creating a single skyrmionium, it is natural to think that the most promising way to produce a SkML is to start from a SkL and induce it to transform into a SkML by appropriate means. Our idea is thus to prepare a SkML by transforming a SkL into a SkML and then let it relax into a minimum energy state.

It is obvious that if the Hamiltonian parameters are the same, the SkML is expected to possess a larger lattice constant than the SkL. Furthermore, if the SkL and SkML share the same lattice constant and other identical Hamiltonian parameters, the SkML will necessarily require a stronger DMI for stabilization. Therefore, in our two methods for creating the SkML, we choose to manually enlarge the lattice constant in the first method, and increase the DMI in the second method to induce the SkL to SkML. Below, we explain our two preparation methods, respectively.

**Method 1**

Generally speaking, in our first method for preparing the SkML, we first fabricate a skyrmion lattice (SkL) with an intentionally enlarged lattice constant by artificially specifying the skyrmion nucleation sites. We then reverse the external magnetic field to promote the nucleation of the SkML on the basis of the preexisting SkL. Subsequently, the second nucleation is again induced at the designated positions, resulting in the formation of the SkML. The cores of skyrmions or skyrmioniums at specific positions can be artificially induced using optical, electrical, or X-ray beams [6-11]. In our simulations, we employ the Monte Carlo method to model spin dynamics under controlled magnetic fields and temperatures, manually specifying certain spins to represent induced cores. The detailed preparation procedure is as follows:

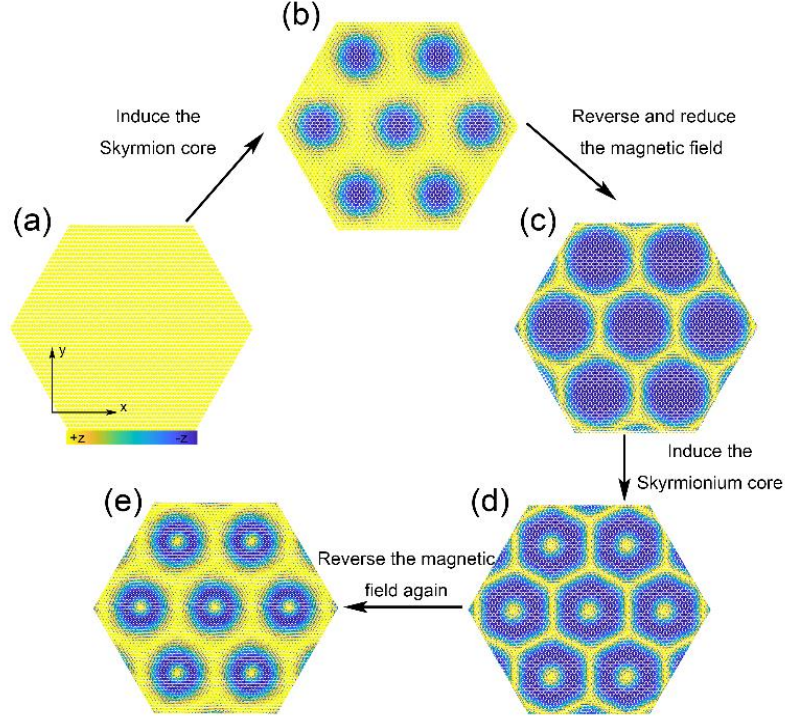

FIG. S1 Experimental preparation of SkML. (a) Ferromagnetic state, where all spins are aligned in the direction of the magnetic field $+h_z$. (b) SkL formed by inducing skyrmion cores at specific positions. (c) Intumescent SkL obtained at low temperatures after reversing and reducing the magnetic field. (d) SkML formed by artificially inducing skyrmionium cores. (e) Final SkML obtained upon reversing the magnetic field again.

We begin with the easily prepared ferromagnetic state of the Hamiltonian Eq. (1) in the main text, as shown in Fig. S1(a). We still take the nearest-neighbor exchange constant $J$ as the unit, and adopt DMI $D=0.41J$, the uniaxial anisotropy $K=0.15J$, and the spin quantum number $S=1$. Under an appropriate magnetic field and temperature ($h_z=0.15J$ for the Zeeman coefficient and $k_BT=0.03J$, where $k_B$ is the Boltzmann constant), the ferromagnetic state naturally transitions into a SkL. However, here we artificially induce skyrmion cores at specific positions by optical, electrical, or X-ray beams. Notably, we increase the distance between two nearest-neighbor skyrmion cores slightly larger compared to naturally formed skyrmion cells to create sufficient space for SkML, as the skyrmionium has a larger radius. Assuming that the nearest-neighbor distance of the triangular lattice is 1, the lattice constant of a typical SkL under these parameters is approximately 16; however, by artificially specifying the nucleation sites, we deliberately increase the lattice constant to 22. After these operations, we obtain SkL shown in Fig. S1(b). Next, we reverse and reduce the magnetic field at low temperature ($h_z=-0.08J$,

$k_BT$=0.03 $J$). After this adjustment, the spin configuration eventually becomes a special SkL under the reversed field [Fig. S1(c)], which we call an intumescent SkL (an expanded SkL) and only exists at low temperatures. Maintaining this low temperature, we then induce skyrmionium cores to form SkML, as shown in Fig. S1(d). However, this SkML in Fig. S1(d) has an edge issue: the skyrmioniums near the lattice edges collapse into skyrmions because the spins at the edge tend to align with the magnetic field. To address this issue, we reverse the magnetic field again and the stable SkML is obtained, as shown in Fig. S1(e).

To illustrate the energy variation during this preparation process, in Fig. S2(a), we compute the average energy per spin at different magnetic fields for all the magnetic states in the experimental preparation process of SkML. The labels "FM", "SkML" and "SkL" represent the ferromagnetic state, skyrmionium lattice state and (intumescent) skyrmion lattice state, respectively. Here the "SkL" contains both the normal SkL and the intumescent SkL state; specifically, it is the intumescent SkL when $h_z$<0 and the normal SkL when $h_z$>0. The black dots a, b, c, d and e represent the five stages in the experimental preparation process correspond to Figs. S1(a)-S1(e). When calculating the average energy per spin in this figure, we directly set the FM, SkML, and SkL as the initial three states, and then perform parameter scanning: change the parameter values and wait for the magnetic structure to stabilize again. Note that during the parameter scanning process, we always keep the evolution of the magnetic structure of the three states at a very low temperature ($k_BT$=0.01$J$) to avoid the thermal fluctuation destroying the corresponding magnetic configurations. After obtaining the magnetic structure corresponding to the certain parameters, the total energy can be calculated by the Hamiltonian (1) in the main text by treating the spins as classical unit vectors. Then, divided by the number of spins, the energy per spin is obtained.

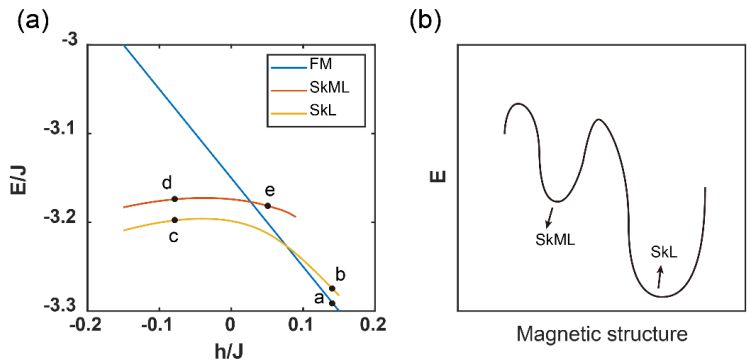

FIG. S2 (a) The average energy per spin in different magnetic fields for the ferromagnetic state, skyrmionium lattice state and skyrmion lattice state, respectively. The black dots a, b, c, d and e represent the five stages in the experimental preparation process correspond to Figs. S1(a)-S1(e). (b) The energy of SkML is higher than SkL, but the SkML is still the local energy minimum state.

These results in Fig. S2(a) also show that under the above parameters, the energy of SkML is larger than the energy of SkL for all magnetic field $h_z$. Moreover, when the magnetic field is positive and small ($0<h_z<0.03J$), the average energy per spin of SkML is less than that of ferromagnetic state.

The experimental preparation stages of SkML can be also described from Fig. S2(a): Starting from the ferromagnetic state, corresponding to point "a" in Fig. S2(a), we first induced the formation of SkL (point "b") via artificial nucleation. Upon further reversal of the magnetic field, the system transitions to point "c", from which the SkML state (point "d") can be readily obtained by artificially inducing the cores of SkML. A final reversal of the magnetic field yields the desired SkML (point "e").

It is worth noting, as shown in Fig. S2(a), that the SkML is indeed not the lowest-energy state. However, the SkML state is a local energy minimum value as Fig. S2(b) shows. Within a certain range of parameters, the SkML can remain stable without collapsing into SkL due to the protection of different topological charges. The phase diagram [Fig. 5(a)] in the main text also shows the stability region of SkML, illustrating the range of magnetic field and temperature under which the SkML can persist.

**Method 2**

Generally speaking, in our second method for creating the SkML, we start from a naturally formed SkL, then create an environment favorable for SkML formation by increasing the DMI, and subsequently induce the SkL to transform into a SkML by adjusting the global temperature and magnetic field. In this fabrication scheme, the most challenging aspect from an experimental perspective is the control of the DMI. However, it has been shown that various techniques for tuning the DMI have already been developed [12-14], and these approaches are capable of achieving the required enhancement of the DMI. Therefore, under ideal conditions, our method is experimentally feasible.

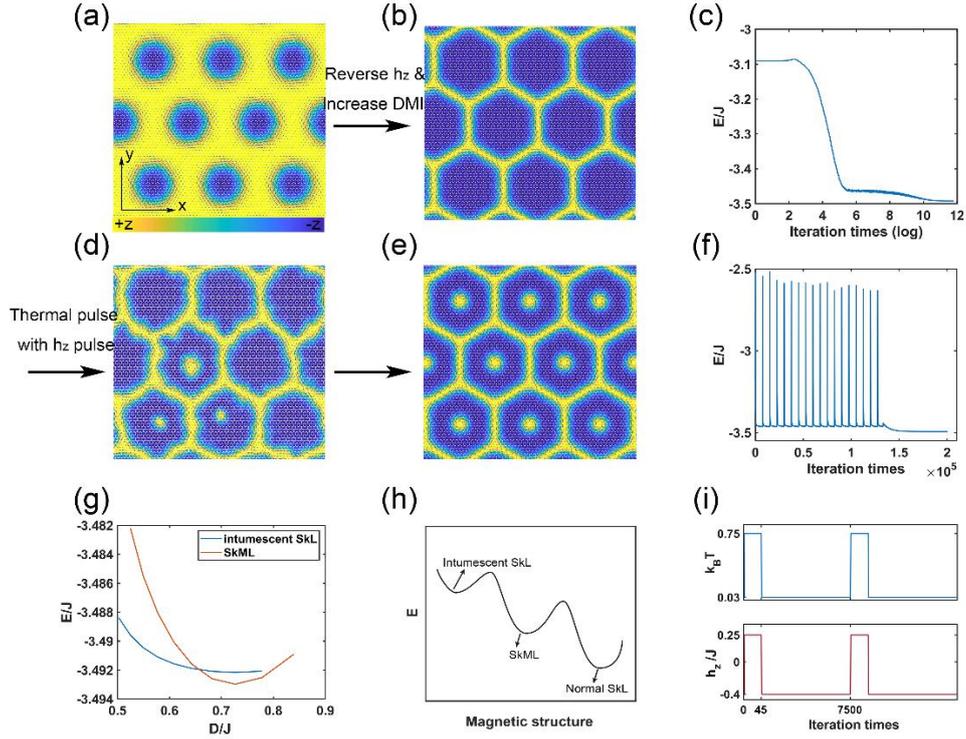

FIG. S3. (a) and (b) are the process of preparing the intumescent SkL in Step 1. (c) The change in average energy per spin during the preparation of the intumescent SkL. (d) and (e) are the process in Step 2 of inducing the SkML from the intumescent SkL. (f) The change in average energy per spin during the SkML preparation in Step 2. (g) The average energy per spin of the intumescent SkL and the SkML as a function of DMI. (h) A schematic of the relative energy levels of the intumescent SkL, the SkML, and the normal SkL. (i) The applied thermal pulses and magnetic field pulses as a function of Monte Carlo iteration steps during the conversion from the intumescent SkL to the SkML.

Our overall procedure for preparing the SkML is illustrated in Fig. S3 and consists of two steps. In our simulated preparation process, we use Monte Carlo simulations on a periodic triangular lattice of size 66×66 spins. We define one iteration as $N$ random spin flips (where $N$ is the total number of spins). Apart from changing global parameters (DMI, temperature, magnetic field) at the start of every step, the spin configuration evolves spontaneously throughout.

**Step 1**: We start from a SkL [Fig. S3(a)] of the Hamiltonian Eq. (1) in the main text. We take the following Hamiltonian parameters: exchange constant $J=1$, uniaxial anisotropy $K=0.15J$, DMI $D=0.3J$, the spin quantum number $S=1$, and the magnetic field $h_z=0.05J$ in $+z$ direction. The nearest-neighbor lattice spacing is set to 1. Under these parameters, the lattice constant of the SkL in the triangular grid is approximately 22. Then, we reverse the magnetic field to $h_z=-0.4J$ and simultaneously, through the DMI-control techniques, increase the DMI to $D=0.73J$. The enhancement of the DMI is introduced to facilitate the subsequent formation of the SkML. After these adjustments, the spin configuration eventually becomes an intumescent SkL under the reversed field [Fig. S3(b)]. The average energy per spin during this evolution is shown in Fig. S3(c), which shows how the energy changes as the normal SkL expands into the intumescent SkL after reversing the field and increasing the DMI.

**Step 2**: This is the most critical step. In this step, we induce the intumescent SkL to transform into the SkML. First, we point out a key consideration: the reason why the intumescent SkL can be induced to become a SkML is that under sufficiently strong DMI, the energy of the SkML becomes close to and even lower than that of the intumescent SkL. As shown in Fig. S3(g), we calculated the average energy per spin for both the SkML and the intumescent SkL as a function of DMI, and the results show that when the DMI exceeds $0.67J$, the energy of the SkML becomes lower than that of the intumescent SkL. This observation suggests an intriguing possibility: a perturbation such as thermal fluctuations may drive the transition from the intumescent SkL [Fig. S3(b)] to the energetically more favorable SkML [Fig. S3(e)]. Note that this does not imply the SkML is the global minimum energy state, because the energy of the SkML is still higher than that of the normal SkL (i.e., the SkL whose core magnetization is opposite to the field); it is only lower than the energy of the intumescent SkL. Although the SkML is not the global energy minimum, it is still a local minimum because it can stably exist within a certain range of magnetic fields and temperatures and will not naturally collapse to other states. In Fig. S3(h), we show a schematic of the relative energies of the intumescent SkL, the SkML, and the normal SkL, indicating that the SkML state is a local energy minimum, which means it can remain stable under certain conditions, but it is not the global energy minimum state. In Fig. 5 of the main text, we demonstrate that a particular SkML remains stable across a finite window of temperature and magnetic field, thereby confirming that the SkML is a locally stable (metastable) state.

In our simulations, we indeed observe that thermal fluctuations can occasionally convert one or two skyrmions into skyrmioniums. However, more often, before such a conversion takes place, the skyrmions are destroyed and neighboring skyrmions merge under the influence of thermal fluctuations. Therefore, we apply thermal pulses instead of continuous heating: the system is briefly heated and then rapidly cooled. In experiment, such global thermal pulses can be implemented, for example, by single ultrafast laser pulses that transiently heat the magnetic film and then allow it to cool via heat diffusion, as in Refs. [6,8], or by short current pulses that provide transient Joule heating, which has been widely used to nucleate skyrmions and skyrmioniums in multilayers [15-17]. To further prevent neighboring skyrmions from merging due to the thermal pulse, we simultaneously apply a magnetic field $+h_z$. This magnetic field favors spin alignment along the +z axis, thereby stabilizing the domain walls between adjacent skyrmions and suppressing their fusion. The global thermal pulses and the +z magnetic field pulses are fully synchronized. When the thermal pulse ends, the +z magnetic field pulse terminates simultaneously, and the magnetic field returns to $h_z=-0.4J$ along the -z direction. The system then waits for the next heat and magnetic field pulses. In addition, since our study is based on Monte Carlo simulations, the thermal pulses and magnetic field pulses are implemented as lasting for 45 Monte Carlo iterations, without an explicit physical time unit. Nevertheless, a reasonable estimate of the relevant time scales for the thermal and magnetic-field pulses can still be obtained through physical considerations. Specifically, the requirement is that the thermal and magnetic field pulses must be applied before the skyrmion shrinks excessively under the external field, so that thermal fluctuations can induce the nucleation of a second core. Previous studies have shown that the field-driven shrinking dynamics of skyrmions occurs on the nanosecond time scale [18,19]. Accordingly, the thermal and magnetic-field pulses considered here should also be on the order of nanoseconds. The generation of thermal pulses via single ultrafast laser pulses or current pulses, as discussed above, is fully capable of reaching nanosecond time scales and is therefore experimentally feasible. The interval between successive thermal (or magnetic field) pulses serves as a cooling period; it only needs to be two orders of magnitude longer than the pulse duration to allow the magnetic texture to sufficiently cool down and re-stabilize. By repeatedly applying thermal pulses in combination with magnetic field pulses as illustrated in Fig. S3(f), the intumescent SkL [Fig. S3(b)] gradually converts into the lower-energy skyrmionium state through an intermediate state [Fig. S3(d)]. Eventually, all intumescent skyrmions are converted into skyrmioniums, ultimately yielding the SkML [Fig. S3(e)]. In the simulations, the efficiency of this transition from the intumescent SkL to the SkML is quite high: we applied only 18 synchronized thermal and magnetic field pulses to completely convert an intumescent SkL containing a 3×3 array of skyrmions (nine skyrmions in total) into a SkML. Finally, we perform a brief annealing to allow the resulting SkML to relax to a local energy minimum, so that the skyrmioniums arrange themselves into a uniform lattice, yielding the final SkML as shown in Fig. S3(e). The energy changes during this step are presented in Fig. S3(f).

The above describes our second method for preparing the SkML. In this approach, whether in the initial preparation of the SkL or in the subsequent induction of the SkL into the SkML, we did not predefine the positions of skyrmions or skyrmioniums; we only adjusted the global parameters (DMI, temperature, and magnetic field) at the start of every step, and the rest of the evolution was natural. Ultimately, this procedure successfully produces a SkML. We also note that accomplishing the transformation from the intumescent SkL to the SkML in Step 2 is certainly not limited to the thermal pulse and magnetic field pulse we used here. We provided just one possible approach. The key point is that when the DMI is within a certain range, the energy of the SkML becomes close to and lower than that of the intumescent SkL. Based on this insight, we devised the above technical route. We hope that our proposed route can provide guidance and inspiration for future experimental realizations. Of course, we believe there are many possible ways to induce a second core inside the SkL. For example, in Ref. [20], a vertical AC magnetic field was used to create a resonant excitation, and in Ref. [2], a spin-polarized current injection was used; both approaches could potentially form a second core from the center of a SkL. Additionally, although our current protocol for preparing the SkML based on SkL is implemented at low temperature, a number of studies have already realized room-temperature SkL experimentally [21-23]. This suggests that applying our SkML preparation route to a room-temperature SkL could enable SkML formation at higher temperatures, thereby substantially relaxing the requirements for SkML fabrication. We consider this a promising direction for further investigation.

**S2 Methods of obtaining $Q$=0 for SkML**

Generally, there are two methods to calculate the topological charge $Q$ [24]:

The first method is the straightforward calculation based on the definition of $Q$. The topological charge $Q$ is given by $Q = \frac{1}{4\pi} \int \boldsymbol{m} \cdot (\partial_x \boldsymbol{m} \times \partial_y \boldsymbol{m}) d^2 r$ [24], where **m** is the unit vector of the local magnetization. Based on the magnetic structure of one skyrmionium unit cell, the topological density $\rho_{top} = \boldsymbol{m} \cdot (\partial_x \boldsymbol{m} \times \partial_y \boldsymbol{m})$ can be numerically calculated, which is also shown in Fig. 3(i) in the main text. Then, by integrating $\rho_{top}$ within the unit cell, the final result for the topological charge $Q$ can be obtained.

Another method is to utilize spherical coordinates [24]. The magnetization $\boldsymbol{m} = (\sin\theta\cos\phi, \sin\theta\sin\phi, \cos\theta)$. In this framework, the topological charge reduces to $Q = -\frac{1}{2}\cos\theta\Big|_{\theta_1=0}^{\theta_2=0} \cdot \frac{1}{2\pi}\phi\Big|_{\phi_1=0}^{\phi_2=2\pi}$. Here, $\theta_1$ and $\theta_2$ are the polar angles of the magnetization at the center and at the edge of the skyrmionium unit cell, respectively. Since in the skyrmionium configuration both the central and peripheral spins point upward ($\theta_1 = \theta_2 = 0$), we have $\cos\theta\Big|_{\theta_1=0}^{\theta_2=0} = 0$, and hence $Q$=0.

**S3 Details of calculating the magnon energy band**

In this supplementary section, we provide a detailed procedure of how to calculate magnon bands in SkML using Holstein-Primakoff (HP) boson theory [25,26] and the para-unitary transformation [27-29].

To describe the quantized magnon in our SkML, we employ the conventional quantum mechanical Holstein Primakoff (HP) boson theory, which represents the magnon system with creation and annihilation operators. Here, we first define the global coordinate system and the local coordinate system. The global coordinate system is a reference frame that is independent of every spin and applies to the whole system. In contrast, the local coordinate system is defined for a specific spin by rotating the original global z-axis to align with the direction of that spin. Each spin possesses its own local coordinate system. We denote the spin operators at site $i$ in the local coordinate system as $\tilde{S}_i^x$, $\tilde{S}_i^y$, $\tilde{S}_i^z$, while the spin operators in the global coordinate system are represented as $S_i^x$, $S_i^y$, $S_i^z$. In the HP theory, the deviation of a spin at lattice site $i$ from its local z-axis is quantified as the boson number $n_i = a_i^\dagger a_i$ at site $i$, and we have

$$\begin{cases} \tilde{S}_i^+ = \sqrt{2S - a_i^\dagger a_i} \, a_i \\ \tilde{S}_i^- = \sqrt{2S - a_i^\dagger a_i} \, a_i^\dagger \\ \tilde{S}_i^z = S - a_i^\dagger a_i \end{cases} \tag{S1}$$

where $\tilde{S}_i^+ = \tilde{S}_i^x + i\tilde{S}_i^y$ and $\tilde{S}_i^- = \tilde{S}_i^x - i\tilde{S}_i^y$; $a_i^\dagger$ is the bosonic creation operator, and $a_i$ is the bosonic annihilation operator. Here we focus on the low-energy excited states of the magnon, such that in the Taylor expansion of Eq. (S1) we may discard the terms higher than quadratic in the magnon operators, so that $\tilde{S}_i^+ = \sqrt{2S} a_i$, $\tilde{S}_i^- = \sqrt{2S} a_i^\dagger$, and we get

$$\begin{cases} \tilde{S}_i^x = \frac{\sqrt{2S}}{2}(a_i + a_i^+) \\ \tilde{S}_i^y = \frac{\sqrt{2S}}{2i}(a_i - a_i^+) \\ \tilde{S}_i^z = S - a_i^\dagger a_i \end{cases} \tag{S2}$$

Next, to express the Hamiltonian in the main text [Eq. (1)] in terms of the creation and annihilation operators in Eq. (S2), we need to rotate the global coordinate system to the local coordinate system for every spin. This allows the spin operators in the global coordinate system to be expressed in terms of the spin operators in the local coordinate system:

$$\begin{cases} S_i^x = \cos\theta\cos\phi\, \tilde{S}_i^x - \sin\phi\, \tilde{S}_i^y + \cos\phi\sin\theta\, \tilde{S}_i^z \\ S_i^y = \cos\theta\sin\phi\, \tilde{S}_i^x + \cos\phi\, \tilde{S}_i^y + \sin\theta\sin\phi\, \tilde{S}_i^z \\ S_i^z = -\sin\theta\, \tilde{S}_i^x + \cos\theta\, \tilde{S}_i^z \end{cases} \tag{S3}$$

where the azimuth angle $0 < \phi < 2\pi$, and elevation angle $0 < \theta < \pi$. Then, by substituting Eqs. (S3) and (S2) into the Hamiltonian Eq. (1) in the main text, the Hamiltonian can be expressed in terms of magnon creation and annihilation operators. Besides, in the correct ground state, the first-order terms of the bosonic operators cancel naturally, leaving only quadratic terms. Upon expressing the Hamiltonian in terms of the magnon creation and annihilation operators, it is convenient to work in the reciprocal space through a Fourier transform $a_{\vec{R},j} = \sum_{\vec{k}} e^{i\vec{R}\cdot\vec{k}} a_{\vec{k},j}$. After some algebra, the magnon Hamiltonian for SkML takes the form:

$$H(\boldsymbol{k}) = \begin{pmatrix} a_k^\dagger & a_{-k} \end{pmatrix} \begin{pmatrix} H(\boldsymbol{k}) & \Delta(\boldsymbol{k}) \\ \Delta^*(-\boldsymbol{k}) & H^*(-\boldsymbol{k}) \end{pmatrix} \begin{pmatrix} a_k \\ a_{-k}^\dagger \end{pmatrix}, \tag{S4}$$

where the vector $\boldsymbol{a_k}$, and the matrices $H(\boldsymbol{k})$ and $\Delta$ each have a dimension of N, the number of spins in the unit cell of SkML. The quadratic bosonic Hamiltonian Eq. (S4) can be diagonalized using the para-unitary transformation [18-20]. Since our focus is on the magnon

dispersion relations and their spatial distribution, specifically, we defined

$$\widetilde{H}(\mathbf{k}) = \sigma_3 \times \begin{pmatrix} H(\mathbf{k}) & \Delta(\mathbf{k}) \\ \Delta^*(-\mathbf{k}) & H^*(-\mathbf{k}) \end{pmatrix}, \text{ with } \sigma_3 = \begin{pmatrix} I & 0 \\ 0 & -I \end{pmatrix}, \quad (S5)$$

where $I$ is the $N$-dimensional identity matrix. The eigenvalues $E(\mathbf{k})$ and eigenvectors $\psi(\mathbf{k})$ of $\widetilde{H}(\mathbf{k})$ can be solved. Here, the eigenvalues $E(\mathbf{k})$ appear in positive and negative pairs. Magnon dispersion is just the positive part of $E(\mathbf{k})$. For the eigenvectors $\psi$, note it should be Boson normalization $\psi^\dagger \sigma_3 \psi = 1$ when calculating the Berry curvature, and it should be probability normalization $\psi^\dagger \psi = 1$ when calculating the magnon density distribution. The magnon density at a given site $i$ is determined by the projection of the eigenvector $\psi$ onto that site: $\rho_i \equiv \langle \psi | a_i^\dagger a_i | \psi \rangle$. Specifically, since each site corresponds to two matrix elements in the 2N-dimensional eigenvector $\psi$, the magnon density at site $i$ is obtained by summing the modulus squared of its two associated matrix elements in $\psi$: $\rho_i = |\psi^i|^2 + |\psi^{N+i}|^2$.

## S4 Details of calculating the band structure of the Haldane model

Based on simplified hexagon lattice shown in Fig. 4(b) in the main text. When we neglect the weakest third-nearest-neighbor coupling $t_3$, and take $t_1=1$, $t_2=0.2$, $\theta_2=-0.4\pi$, the Hamiltonian in the momentum space can be expressed as

$$H(\mathbf{k}) = \begin{pmatrix} H_{11} & H_{12} \\ H_{21} & H_{22} \end{pmatrix}, \quad (S6)$$

where

$$H_{11} = t_2 e^{i\theta_2} e^{ik_x\sqrt{3}a} + t_2 e^{i\theta_2} e^{-\frac{ik_x\sqrt{3}a}{2}-\frac{ik_y 3a}{2}} + t_2 e^{i\theta_2} e^{-\frac{ik_x\sqrt{3}a}{2}+\frac{ik_y 3a}{2}} + c.c.$$

$$H_{12} = t_1 e^{-ik_y a} + t_1 e^{-ik_x\sqrt{3}a/2 + ik_y a/2} + t_1 e^{ik_x\sqrt{3}a/2 + ik_y a/2}$$

$$H_{21} = H_{12}^*$$

$$H_{22} = t_2 e^{i\theta_2} e^{-ik_x\sqrt{3}a} + t_2 e^{i\theta_2} e^{ik_x\sqrt{3}a/2 + ik_y 3a/2} + t_2 e^{i\theta_2} e^{ik_x\sqrt{3}a/2 - ik_y 3a/2} + c.c. \quad (S7)$$

Here, "c.c." denotes the complex conjugation of the preceding mathematical expression. Based on this Hamiltonian matrix, the eigenvalues are solved, as the band structure shown in Fig. 4(c) in the main text, where the momenta are selected along the high-symmetry points as the inset shows.

## S5 Analysis of thermal Hall conductivity and energy band structures

In this section of the supplementary material, we provide an explanation for the three characteristics of thermal Hall conductivity discussed in the main text by analyzing the energy band structures, Chern numbers, and Berry curvature of different magnetic states.

The thermal Hall conductivity is calculated in the main text at three temperatures: $k_BT$=0.04$J$, 0.07$J$, and 0.1$J$. Note that the magnetic structures at each temperature are all relaxed to their minimum energy at zero temperature before evaluating the thermal Hall conductivity. Consequently, the magnetic structures in the same Zeeman coupling coefficient and same SkML/SkL phase at these three temperatures are nearly identical when minimized at zero temperature. The effect of temperature in the same Zeeman coupling coefficient and same SkML/SkL phase is considered only reflected in the occupation of magnons. So the band structure can be considered as only dependent on the Zeeman coupling for the SkML/SkL phase. And the energy bands of corresponding magnetic structures for different Zeeman coupling coefficients ($h_z$=0, 0.03, 0.05, 0.09, 0.12, 0.16$J$) can be calculated as shown in Figs. S6(a)-S6(f). Figs. S6(a)-S6(c) correspond to SkML phase, while Figs. S6(d)-S6(f) correspond to SkL phase. The momenta are selected along the high-symmetry points of the Brillouin zone, as illustrated in the bottom-right corner. The color of the bands in the figure represents the magnitude of the Berry curvature at the corresponding momentum-space points. Since the contribution of magnons to the thermal Hall conductivity is primarily determined by the lower-energy bands in low-energy excitations, we present only the lowest fourteen energy bands, which are sufficient to provide a clear explanation. The Chern numbers of the fourteen energy bands for each diagram in Fig. S6, listed from lowest to highest, are as follows:

For Fig. S6(a): $C$ = (0, -1, 3, 1, -1, -2, 0, 1, -1, 0, 3, 0, -5, -2), SkML phase;

For Fig. S6(b): $C$ = (0, -1, 3, 1, -3, 0, 0, 1, -1, -1, 4, -3, 1, -1), SkML phase;

For Fig. S6(c): $C$ = (0, -1, 3, 1, -3, 0, -1, 2, -1, -3, 6, 2, -5, -3), SkML phase;

For Fig. S6(d): $C$ = (0, 0, 0, 0, 0, 3, 1, 0, 1, 0, 2, 1, 1), SkL phase;

For Fig. S6(e): $C$ = (0, 0, 0, 0, 1, 2, 1, -1, 2, 2, 1, 1, -1, 2) , SkL phase;

For Fig. S6(f): $C$ = (0, 0, 0, 0, 3, 2, -2, 2, -2, 5, -1, 2, -3, 0) , SkL phase.

Based on these energy band structures in Fig. S6, we now explain the three characteristics of the thermal Hall conductivity in turn.

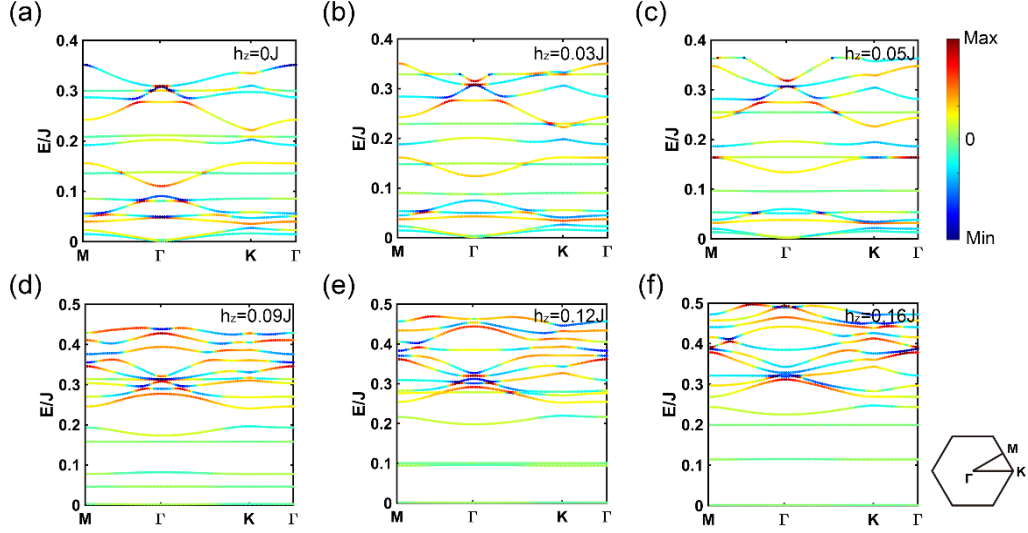

FIG. S6. Band structure at different magnetic field strengths. (a)-(c) correspond to SkML phases, and (d)-(f) correspond to SkL phases. The momenta are selected along high-symmetry points of the Brillouin zone (bottom right corner). The color of the bands in the figure represents the magnitude of the Berry curvature at the corresponding momenta.

First feature: In both SkML and SkL phases, the thermal Hall conductivity decreases with increasing magnetic field. Let's discuss SkL phase first. As shown in Figs. S6(d)-S6(f), the energy bands with positive Berry curvature shift upward with increasing magnetic field in SkL phases, resulting in an overall increasing energy of these bands. Since magnons tend to occupy lower-energy states, the upward shift of the energy bands reduces the number of magnons that can be excited, thereby decreasing their contribution to the thermal Hall conductivity. In SkML phase as shown in Figs. S6(a)-S6(c), the phenomenon of energy bands with positive Berry curvature shifting upward also happens but not very obvious. However, the phenomenon of energy bands with negative Berry curvature shifting downward is obvious in Figs. S6(a)-S6(c), resulting in the same effect.

Second feature: The overall thermal Hall conductivity in SkML phase is lower than that in SkL phase. This is because, in the lower energy bands (e.g., the first fourteen bands shown in Fig. S6), the Chern numbers of SkL phase are predominantly positive, whereas the number of lower-energy bands with positive Chern numbers in SkML phase is significantly smaller. This is also evident from the color distribution of the energy bands in Fig. S6, where regions with positive Berry curvature are more prevalent in Figs. S6(d)-S6(f) compared to Figs. S6(a)-S6(c).

Third feature: The thermal Hall conductivity of SkL phase decreases sharply as the temperature decreases, while the thermal Hall conductivity of SkML phase decreases more moderately. This is because, at lower temperatures (e.g., $k_BT$=0.03$J$), magnons occupy lower-energy states. In SkL phase (Figs. S6(d)-S6(f)), the Chern numbers and Berry curvatures of several lowest-energy bands (e.g., those with energy less than 0.1$J$) are very small, leading to a significant reduction in the contribution of Berry curvature to the thermal Hall conductivity. As a result, the thermal Hall conductivity in SkL phase can even become lower than in SkML phase when the temperature is very low. In contrast, in SkML phase [Figs. S6(a)-S6(c)], several lowest-energy bands (e.g., those with energy less than 0.1$J$) with positive Chern numbers or Berry curvatures remain, causing only a slight decrease in the thermal Hall conductivity as the temperature lowers, with a much smaller reduction compared to SkL phase.